\documentclass[aps,showpacs,twocolumn,superscriptaddress]{revtex4}
\usepackage{graphicx}
\usepackage{bbm}
\usepackage{amssymb}
\usepackage{epstopdf}
\usepackage{color, amsmath, bm}
\usepackage[hidelinks]{hyperref}
\usepackage{natbib}
\newcommand{\conmutador}[2]{\left[#1,#2\right]}

\newcommand{\ci}[1]{\text{cosint}\left(#1\right)}
\newcommand{\si}[1]{\text{sinint}\left(#1\right)}

\RequirePackage{amsmath}
\usepackage{amsfonts}   
\usepackage{physics}

\begin{document}

\title{Boundary-induced effect encoded in the corrections to the geometric phase acquired by a bipartite two-level system}

\author{Ludmila Viotti, Fernando C. Lombardo and Paula I. Villar}
\affiliation{Departamento de F\'\i sica {\it Juan Jos\'e
Giambiagi}, FCEyN UBA and IFIBA CONICET-UBA, Facultad de Ciencias Exactas y Naturales,
Ciudad Universitaria, Pabell\' on I, 1428 Buenos Aires, Argentina.}
\date{\today}                                           

\begin{abstract}  
We present a bipartite two-level system coupled to electromagnetic quantum vacuum fluctuations through a general dipolar coupling. We derive the master equation in the framework of open quantum system, assuming the environment composed of (i) solely vacuum fluctuations and (ii) the vacuum fluctuations and a conducting plate located at a fixed distance from the bipartite system.  For both cases considered, we study the dynamics of the bipartite system and the temporal evolution of the concurrence of an initial entangled bipartite state. We further analyse the generation of entanglement due to the vacuum structure. Finally, we study the different induced contributions to the correction of the unitary geometric phase of a bipartite quantum state  so as to explore the possibility of future experimental setups by exploring the influence of boundaries conditions in vacuum.
\end{abstract}

\maketitle

\section{Introduction}

The global phase acquired by a quantum system related to its dynamical evolution  contains a gauge-invariant component, namely, the geometric phase (GP) which depends only on the geometry of the path traversed by the system during the quantum evolution \cite{Berry, Pancharatnam}. Since their seminal works, great progress has been achieved in this field. As an important evolvement, the application of the geometric phase has been proposed in many fields, such as the geometric quantum computation. Due to its global properties, the geometric phase is propitious to construct fault tolerant quantum gates. In this line of work, many physical systems have been investigated to realize geometric quantum computation, such as NMR (Nuclear Magnetic Resonance) \cite{jones} , Josephson junction \cite{faoro}, Ion trap \cite{duan} and semiconductor quantum dots \cite{solinas}. The quantum computation scheme for the GP has been proposed based on the Abelian or non-Abelian geometric concepts, and the GP has been shown to be robust against faults in the presence of some kind of external noise due to the geometric nature of Berry phase \cite{zanardi,Sjoqvist_2012}.

Due to the fact that a real quantum system unavoidably interacts with its environment and undergoes decoherence, much attention has been raised by studies on the geometric phase in open quantum systems under non-unitary dynamics as it has been shown that the interactions play an important role for the realization of some specific operations.  Since the gates operate slowly compared to the dynamical time scale, they become vulnerable to open system effects and parameters  fluctuations that may lead to a loss of coherence. Consequently, the study of the geometric phase was soon extended to open quantum systems and many authors have analyzed the robustness GP under the influence of a wide variety of external environments, by the use of different approaches 
 \cite{carollo,carollo_2,carollo_3,chiara,tong1,tong2,Yi_2005,Rezakhani,Yi_2006,042311,707713,VILLAR2009206,Yin_2009,Yin_2010,Chen_2010,Wu_2010,052121,032338,pra,erikreview,
 Luo_2018,Cai_2019,carollo_4}. The GP is a promising building block for noise-resilient quantum operations. Lately, the GP has been observed in a variety of superconducting systems \cite{berger,leek} and,  in \cite{prl}, the geometric phase of an open system undergoing nonunitary evolution has been measured using a NMR quantum simulator.
In this context, GPs have become a fruitful avenue of investigation to infer features of the quantum system due to their topological properties and close connection with gauge theories of quantum fields.  Since the phase depends only on the system  path in parameter space, particularly the flux of some gauge field enclosed by that path. Then, in \cite{guridi_1}  the GP is said to encode information about the number of particles in the field for pure field states. In particular, for initial squeezed states, the phase also depends on the squeezing strength \cite{guridi_2}. If the field is in a thermal state, the GP encodes information about its temperature, and so it has been proposed  to measure the Unruh effect at low accelerations \cite{martin-martinez}. It has further been proposed as a high-precision thermometer by considering the atomic interference of two atoms interacting with a known hot source and an unknown temperature cold cavity \cite{martin_martinez_2}. In addition, in \cite{fasechinos}, authors suggest a possible way of detecting vacuum fluctuations in experiments involving the geometric phase. Recently, it was found a proper scenario to indirectly detect the non-contact quantum friction \cite{pendry97} by measuring the GP acquired by a particle moving in front of a dielectric plate, where they have even proposed an experimental setup which determined the feasibility of the experiment  as for  tracking traces of quantum friction through the study of decoherence effects and the correction of the unitary geometric phase on a two-level system \cite{npjqi}.  Therein, it has been shown that the GP can be used to infer quantum properties of the systems with the emergence of new technologies.
 
Quantum entanglement is another important concept in quantum physics, playing a central role in many novel quantum technologies \cite{bennett1,bennett2}. The understanding of GPs for entangled states is particularly relevant due to potential applications in holonomic quantum computation with spin systems, which provide a plausible design of a solid-state quantum computer. Quantum and classical correlations alike always decay as a result of noisy backgrounds and decorrelating agents that reside in ambient environments \cite{zurek}, so the degradation of entanglement shared by two or more parties is unavoidable. It has been said that two initially entangled atoms may get completely disentangled within a finite time, which is known as entanglement sudden death \cite{book_nielsen,yu}. 
Likewise,  a common bath can also provide indirect interactions among independent atoms, leading to entanglement sudden birth \cite{maniscalco}. Thus, the definition and modeling of the environment becomes relevant. Particularly, vacuum quantum fluctuations are a type of environment that can not be turned off. The natural coupling of a neutral atom to the electromagnetic quantum fluctuations is through the dipolar interaction. It is well known that electromagnetic quantum fluctuations are modified by the presence of boundaries and the resulting distortions are known as observable effects such as the Lamb shift \cite{lamb} and the Casimir effect \cite{casimir}. The GP has been studied for bipartite systems in \cite{042111}, showing that initially maximally entangled (MES) state acquired a ``robust" GP \cite{OXMANposta}.  

In this context, questions naturally arise as to what extend the presence of boundary conditions modify the geometric phase and the entanglement on a bipartite system. As has been indicated above, the geometric phase is modified by the presence of quantum vacuum fluctuations. As the electromagnetic quantum vacuum fluctuations are modified by the presence of boundaries, we shall expect to obtain traces of that non trivial vacuum modification in the corrections of the geometric phase of the bipartite system.  As the GP is known to be less corrected in a MES state  \cite{042111}, we shall extend the study of the corrections to the GP to a bipartite state coupled to vacuum fluctuations.   The ultimate goal is to study the bipartite dynamics, concurrence, and geometric phase in the presence of the electromagnetic quantum vacuum fluctuations and compare them with the those obtained when boundary conditions are changed (and therefore quantum vacuum structure is modified). Furthermore, it may be useful to know if the influence of these boundaries can be exploited for an experimental test on the GP as so to infer quantum properties of the systems \cite{OXMANposta}. This would help to define future experimental setups, where the geometric phase would be used as a tool to sense traces of observables consequences of quantum vacuum fluctuations \cite{npjqi}.

This paper is structured as follows. In Section \ref{system_section}, we present the model studied and derive the master quantum equation so as to describe the dynamics  composed of two two-level systems coupled to electromagnetic quantum vacuum fluctuations through a dipolar coupling. We derive the environmental kernels for two situations: (i) free space vacuum fluctuations and (ii) a conducting plate located at a fixed distance of the bipartite quantum system.  In Sec. \ref{dinamica} we study the dynamics of an initially entangled state under the presence of quantum vacuum fluctuations, focusing on the effect when boundaries are considered. In Sec. \ref{concurrence}, we extend the analysis to the entanglement dynamics of the bipartite system, by considering how the initial entanglement of the quantum system is influenced by the environment. Further, we study if the presence of the environment can generate entanglement in an initial separate state and the functional dependence of the concurrence upon the distance to the conducting plate and the distance among particles. In Sec.\ref{phase} we compute the GP acquired by the bipartite system in either situation considered: (i) free space and (ii) in presence of a conducting plate in order to compare both situations and see if the presence of boundaries conditions can be exploited for future measurements the GP. Finally, in Sec. \ref{conclusiones}, we summarize the results and present  conclusions.

\section{The system}\label{system_section}
\begin{figure}[h]
\centering
\includegraphics[width=.5\columnwidth]{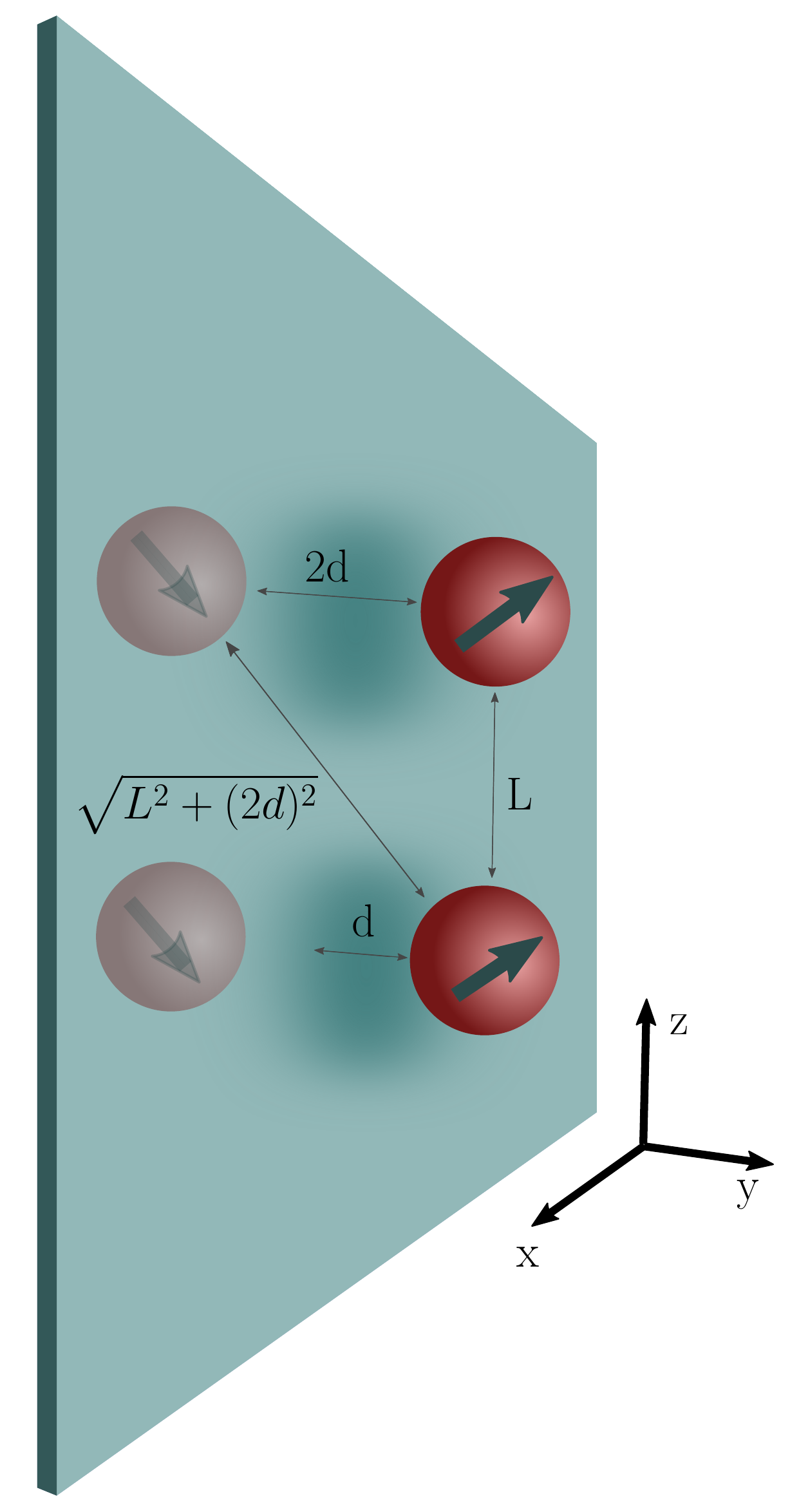}
\caption{\label{esquema} A scheme of the system under consideration, where the two-level atoms are at a fixed distance $d$ from a perfectly conducting plate and separated by a distance $L$. The electromagnetic field in presence of a conducting plate, can be thought of as that produced by the atoms and two image dipoles.}
\end{figure} 

We shall consider a bipartite system consisting in two, originally non-interacting, two-level systems. Both two level systems are coupled to the electromagnetic field in its vacuum state (and all in turn can be  in presence of a perfectly conducting plate). The model, which is schematically represented in Fig. \ref{esquema}, can be mathematically described by a Hamiltonian 
\begin{equation}
    H=H_{\text{s}}+H_{\text{em}}+H_{\text{int}},
\end{equation}
where the first two terms $H_{\text{s}}$ and $H_{\text{em}}$ are the Hamiltonian of the free bipartite system and the electromagnetic field in presence of a perfectly conducting plate respectively, and $H_{\text{int}}$ is the interaction Hamiltonian between the system and the vacuum field. The free system Hamiltonian can be written as
\begin{equation}
    H_{\text{s}}=\frac{\hbar}{2}\omega_0^1\sigma_3 \otimes \mathbbm{1} +\frac{\hbar}{2}\omega_0^2 \mathbbm{1} \otimes \sigma_3,
\end{equation}
while the dipolar interaction Hamiltonian (between each atom and the electromagnetic field) can be expressed as $H_{\text{int}}=H_{\text{int}}^1+H_{\text{int}}^2$, with
\begin{align}\nonumber
    H_{\text{int}}^i=&- \hat{\mathbf{\mu}}^i\cdot\hat{\mathbf{E}}(t,x^i)\\ \nonumber
    =&-e\sum_{\lambda=1}^2\int\frac{d^3k}{(2\pi)^3}\sqrt{2\pi\hbar\omega_{\mathbf{k}}}\left(\mathbf{r}^i_+\sigma^i_++\mathbf{r}^i_-\sigma^i_-\right)\\
    &\hspace{1.3cm}\times \left[a_{\lambda\mathbf{k}}A_{\lambda\mathbf{k}}(\mathbf{x^i})-a^{\dagger}_{\lambda\mathbf{k}}A^*_{\lambda\mathbf{k}}(\mathbf{x^i})\right]\mathbf{\epsilon}_{\lambda\mathbf{k}},
\end{align}
where $i$ implies a label on each particle, $\hat{\mathbf{\mu}}^i=e\left(\mathbf{r}^i_+\sigma^i_++\mathbf{r}^i_-\sigma^i_-\right)$ is the electric dipole moment operator of atom i, $e$ is the electron charge and $\mathbf{r}^i_{\pm}$ the dipole orientations; each particle located at a spatial position $\mathbf{x}$. We are assuming $\mathbf{r}$ to be a classical parameter with the operatorial character of $\hat{\mathbf{\mu}}$ contained in $\sigma_\pm^i$.
We have also used the lower order multipolar Hamiltonian, with $a_{\lambda\mathbf{k}}$ the bosonic destruction operator of momentum $k$ and polarization $\lambda$ and the mode functions $A_{\lambda\mathbf{k}}(\mathbf{x})$ of the electromagnetic field, which should be chosen as to satisfy the boundary conditions \cite{book_milonni}.
 The  free space mode functions  should satisfy the boundary condition,
\begin{equation}
    A_{\lambda\mathbf{k}}(\mathbf{x}^i)=e^{-i\mathbf{k}\mathbf{x}^i},
\end{equation}
while the functions considering the plate, in the region $y\geq 0$ are
\begin{align}
    A_{1\mathbf{k}}(\mathbf{x}^i)&=\sqrt{2}\;(\check{k}_{\parallel}\times\check{k}_{\perp})e^{i\mathbf{k}_{\parallel}\mathbf{x}^i_{\parallel}}\\\nonumber
    A_{2\mathbf{k}}(\mathbf{x}^i)&=\frac{\sqrt{2}}{k}\left[k_{\parallel}\cos(\mathbf{k}_{\perp}\mathbf{x}^i_{\perp})\check{k}_{\perp}-ik_{\perp}\sin(\mathbf{k}_{\perp}\mathbf{x}^i_{\perp})\check{k}_{\parallel}\right],
\end{align}
with $\mathbf{x}^1=(0,d,L)$, $\mathbf{x}^2=(0,d,0)$,  $\mathbf{k}_{\parallel}=(k_1,0,k_3)$ and $\mathbf{k}_{\perp}=k_2\check{y}$. We're applying the inverted hat to denote classical unit vectors. \\

As we want to know the dynamics of the bipartite system at all times, we shall derive a master quantum equation by applying the formalism of open quantum systems \cite{petruccione}. The state of the system, which is represented by the reduced density matrix for the bipartite system, satisfies a master equation. 
If we perform a power expansion, up to second order in series of the coupling between the bipartite system and the environment (composed by the electromagnetic field plus the mirror), this equation is given by 
\begin{align}
    \dot{\rho}_\text{s}(t)=&\frac{1}{i\hbar}\Tr_\epsilon\left(\conmutador{V(t)}{\rho_{\text{s}}(t)\otimes \rho_\epsilon}  \right) \label{eqlarga}\\ \nonumber -&\frac{1}{\hbar^2}\int_0^t dt' \Tr_\epsilon{\left(\conmutador{V(t)}{\conmutador{V(t')}{\rho_{\text{s}}(t)\otimes \rho_\epsilon}}\right)}\\ \nonumber
    +&\frac{1}{\hbar^2}\int_0^tdt' \Tr_\epsilon\left(\conmutador{V(t)}{\Tr_\epsilon{\left(\conmutador{V(t')}{\rho_{\text{s}}(t)\otimes \rho_\epsilon}\right)}\otimes \rho_\epsilon}\right)
\end{align}
where  $V(t)=U_0^\dagger \, H_{\text{int}} \, U_0$  is the interaction Hamiltonian in the interaction picture and $U_0=e^{-\frac{i}{\hbar}(H_\text{s}+H_\epsilon)t}$ is the operator representing the evolution produced by the free theories. By taking the initial state of the electromagnetic field to be the vacuum state, both the first and the third terms in Eq. (\ref{eqlarga}) vanish, leaving a simplified equation governing the dynamics of the bipartite system: 
\begin{equation}
    \dot{\rho}_\text{s}(t)=-\frac{1}{\hbar^2}\int_0^t dt' \Tr_\epsilon{\left(\conmutador{V(t)}{\conmutador{V(t')}{\rho_{\text{s}}(t)\otimes \rho_\epsilon}}\right)}.
    \label{eqcorta}
\end{equation}

The interaction $V(t)$ present in Eq. (\ref{eqcorta}) can be found through a simple calculation involving Hausdorff formula, resulting in
\begin{eqnarray}\nonumber
    V^i(t)&=&-e\sum_{\lambda=1}^2\int\frac{d^3k}{(2\pi)^3}\sqrt{2\pi\hbar\omega_{\mathbf{k}}}\\ \nonumber
    &\times& \left[\mathbf{r}_+^i(\sigma_+^i\otimes a_{\lambda\mathbf{k}}) \,A_{\lambda\mathbf{k}}(\mathbf{x}^i) \,e^{-i(\omega_{\mathbf{k}}-\omega_0^i)t} +\right.\\ \nonumber
    & &\hspace{.2cm} \mathbf{r}_+^i(\sigma_+^i\otimes a_{\lambda\mathbf{k}}^\dagger) \,A^*_{\lambda\mathbf{k}}(\mathbf{x}^i) \,e^{i(\omega_{\mathbf{k}}+\omega_0^i)t} +\\\nonumber
    & &\hspace{.2cm} \mathbf{r}_-^i(\sigma_-^i\otimes a_{\lambda\mathbf{k}}) \,A_{\lambda\mathbf{k}}(\mathbf{x}^i) \,e^{-i(\omega_{\mathbf{k}}+\omega_0^i)t} +\\
   & & \hspace{.15cm} \left.\mathbf{r}_-^i(\sigma_-^i\otimes a^\dagger_{\lambda\mathbf{k}})\,A^*_{\lambda\mathbf{k}}(\mathbf{x}^i)\,e^{i(\omega_{\mathbf{k}}-\omega_0^i)t} \right]\mathbf{\epsilon}_{\lambda\mathbf{k}}.
    \label{vdet}
\end{eqnarray}

With Eqs. (\ref{eqcorta}) and (\ref{vdet}) as the starting point and after some algebra, the master equation governing the dynamics of the reduced density matrix for the bipartite system can be written, in the secular approximation:
\begin{align}
    &\dot{\rho}_s(t)=\frac{1}{i\hbar}[H_s,\rho_s] \\ \nonumber
    &- \sum_{i,j=1}^2\left[\int_0^t dt_1K^{ij}_{-}(t')\left([\sigma_+^i,\{\sigma_-^j,\rho_s\}]+[\sigma_+^i,[\sigma_-^j,\rho_s]] \right)+ \right.\\ \nonumber
    &\left.\int_0^t dt_1K^{ij}_{+}(t') \left([\sigma_-^i,\{\sigma_+^j,\rho_s\}]+[\sigma_-^i,[\sigma_+^j,\rho_s]] \right) + \text{h.c.}\right], 
\end{align}
valid as long as the relaxation time $\tau_{R}$ and all evolution timescales of the system $\tau_S\sim 1/\omega_0^i$ satisfy the relation $\tau_R\gg\tau_S$ \cite{petruccione}. $K^{ij}_{\pm}(t')$ are kernels containing all the information about the effect of the electromagnetic field on the system.
\begin{align}
     K^{ij}_{\pm}(t')&=2\pi\frac{e^2}{\hbar}|\mathbf{r}|^2\sum_{m,n=1}^3 r^i_m r^j_n\int \frac{d^3\mathbf{k}}{(2\pi)^3}\;\omega_{\mathbf{k}}
    \\[.8em]\nonumber
    &\times  \left(\sum_{\lambda}A_{\lambda\mathbf{k}}(\mathbf{x}^i)A^*_{\lambda\mathbf{k}}(\mathbf{x}^j) \right)\,e^{-i(\omega_{\mathbf{k}}\pm\omega_0)(t-t')}.
\end{align}

In order to obtain these kernels we have made some further assumptions. We are assuming that both atoms have the same natural frequency $\omega_0$ and that they have the same dipolar moment magnitude $|\mathbf{r}|$ allowing for the moment the directions to be different. These directions are represented by two unit vectors $\check{r}^i$ whose components are $r^i_m$. 
The master equation can be written in a more suggestive form
\begin{align}
    &\dot{\rho}_s(t)=\frac{1}{i\hbar}[H_s,\rho_s]-i\sum_{i=1}^2 c_{ii}(t)[\sigma_+^i\sigma_-^i,\rho_s] \\ \nonumber
    &-i\sum_{i\neq j}c_{ij}(t)\left[\sigma_+^i\sigma_-^j,\rho_s\right]\\\nonumber
    &-\sum_{i=1}^2 a_{ii}(t)\left(\sigma_+^i\sigma_-^i\rho_s +\rho_s\sigma_+^i\sigma_-^i-2\sigma_-^i\rho_s\sigma_+^i\right) \\ \nonumber
    &-\sum_{i\neq j} a_{ij}(t)\left(\sigma_+^i\sigma_-^j\rho_s + \rho_s\sigma_+^j\sigma_-^i - \sigma_-^j\rho_s\sigma_+^i  - \sigma_-^i\rho_s\sigma_+^j \right).\\\nonumber
    \label{eq_diff_coef}
\end{align}
The first term corresponds to the free evolution of the system (unitary evolution) while the second and third terms  are  the frequency renormalization and an effective interaction terms. The last two terms in this equation are responsible for dissipation and fluctuations (noise) effects. In the previous equation we have used that $\sigma_\pm = \sigma_x \pm i \sigma_y$.  The information about the environment (with or without boundaries) is thus encoded in the kernels $a_{ij}(t)$ and $c_{ij}(t)$, which in the markovian approximation can be computed by direct integration as 

\begin{eqnarray}
a_{ij} &=& \Re\int^\infty_{0}dt'\;K^{ij}(t'), \\
c_{ij} &=_{i\neq j} & \Im\int^\infty_{0}dt'\;\left(K^{ij}_{+}(t')+K^{ij}_{-}(t')\right),\\
c_{ii} &=& \Im\int^\infty_{0}dt'\;\left(K^{ii}_{+}(t')-K^{ii}_{-}(t')\right), \label{definicioncij}
\end{eqnarray}
It is easy to note in Eqs.(\ref{definicioncij}) that the coefficient $c_{ii}$ represents a frequency renormalization,  while the term with $c_{ij}$ represents an effective interaction among the components of the bipartite (the renormalization of the coupling constant between the two-level systems considered \cite{Hu}). The coefficient $a_{ij}, i\neq j$ is usually referred to as collective damping. 
It is important to note, that the kernels will have different expressions whether we are considering the bipartite system either in free space (solely coupled to quantum vacuum fluctuations) or the presence of a perfectly reflecting boundary in quantum vacuum. The explicit expression for the environment $a_{ij}(t)$ and $c_{ij}(t)$ kernels can be found in Appendix \ref{appendixa}, and will remain valid as long as markovian approximation does. This is, as long as the relaxation time $\tau_R$ and the vacuum field correlation time $\tau_E$ (the characteristic width of the environment correlation functions) satisfy the condition $\tau_R\gg\tau_E$, which imposes the conditions, in natural units, $L\omega_0\gtrsim 1$ and $2 d \omega_0\gtrsim 1$ \cite{maniscalco3}.\\

Finally, by assuming that $K^{12}(t)=K^{21}(t)$ and $K^{11}(t)=K^{22}(t)$, the master equation derived can be analytically solved for a general initial state of the bipartite qubit system of the form 
\begin{equation}
\ket{\psi}=\alpha\ket{11}+\beta\ket{10}+\gamma\ket{01}+\sigma\ket{00}.
\label{initial_state_general}
\end{equation}
The resulting matrix elements composing the reduced density matrix $\rho_s(t)$ are explicitly given in Appendix \ref{appendixb}.

\section{System's open dynamics: in free space and with a reflecting boundary}\label{dinamica}

We shall start studying the dynamical properties of the system, whether the bipartite system is in free space or at a fixed distance of a reflecting plate, located at $y=0$.  In both cases, the atoms are separated a distance L in $\check z$ direction. If we consider the presence of a plate, we shall also assume the atoms to be fixed at a distance $d$ from the perfectly conducting plate in the $\check{y}$ direction.
As we are interesting in the effect of the vacuum fluctuations (either dressed or undressed) on an initial state of the quantum system, 
 we shall define an initial  bipartite state of the form
\begin{equation}
    \ket{\psi(0)}=\sqrt{p}\ket{11}+\sqrt{1-p}\ket{00}, 
    \label{estado_1}
\end{equation}
in the $\{\ket{00},\ket{01},\ket{10},\ket{11}\}$ basis. In Eq.(\ref{estado_1}), $p$ determines the degree of entanglement being $\ket{0}$ and $\ket{1}$ eigenstates of the Pauli operator $\sigma_z$ (of each two level system). It is easy to note that $p=1/2$ accounts for a maximally entangled state (MES). As has been indicated in \cite{maniscalco}, there is no entanglement sudden death for any generic state with maximum one excitation. Analogously, entanglement can be smoothly generated but it cannot suddenly appear.  That is the reason we shall limit to study this type of Bell-like state. Taking this particular initial state, the reduced density matrix of the bipartite system assumes the simplified form 
 \begin{equation*}
    \rho(t)=\left(\begin{array}{cccc}
        \rho_{11}(t) &0 &0 &\rho_{14}(t) \\
        0 & \rho_{22}(t)& \rho_{23}(t) &0 \\
        0 &\rho_{32}(t) &\rho_{33}(t) &0 \\
       \rho_{41}(t) &0 &0 &\rho_{44}(t)
    \end{array}\right).
\end{equation*}
Full expressions of the components of the reduced density matrix can be found in Appendix \ref{appendixb}.
As decoherence is the dynamical suppression of quantum coherences, the off-diagonal elements of the reduced density matrix are a good measure of how the environment affects the dynamics of the system. In our study, the only off-diagonal elements of interest are $\rho_{23}(t)$ and $\rho_{41}(t)$, being the remaining off-diagonal elements either zero, or determined by these two. In what follows, we will be working in natural units $c=\hbar=1$ and, in this units, we will take the dipolar coupling to be of the order of Bohr radius $|\mathbf{r}|\sim a_0$ and the natural frequency to be of the order of the hydrogen ground state energy $\omega_0\sim E_0 =10^{6} 1/m$ leading to $\gamma_0 =1$ and allowing for expansions in powers of  $\gamma_0/\omega_0$ and satisfying all the approximations that have been done. 

\begin{figure}[ht]
\centering
\includegraphics[width=.8\columnwidth]{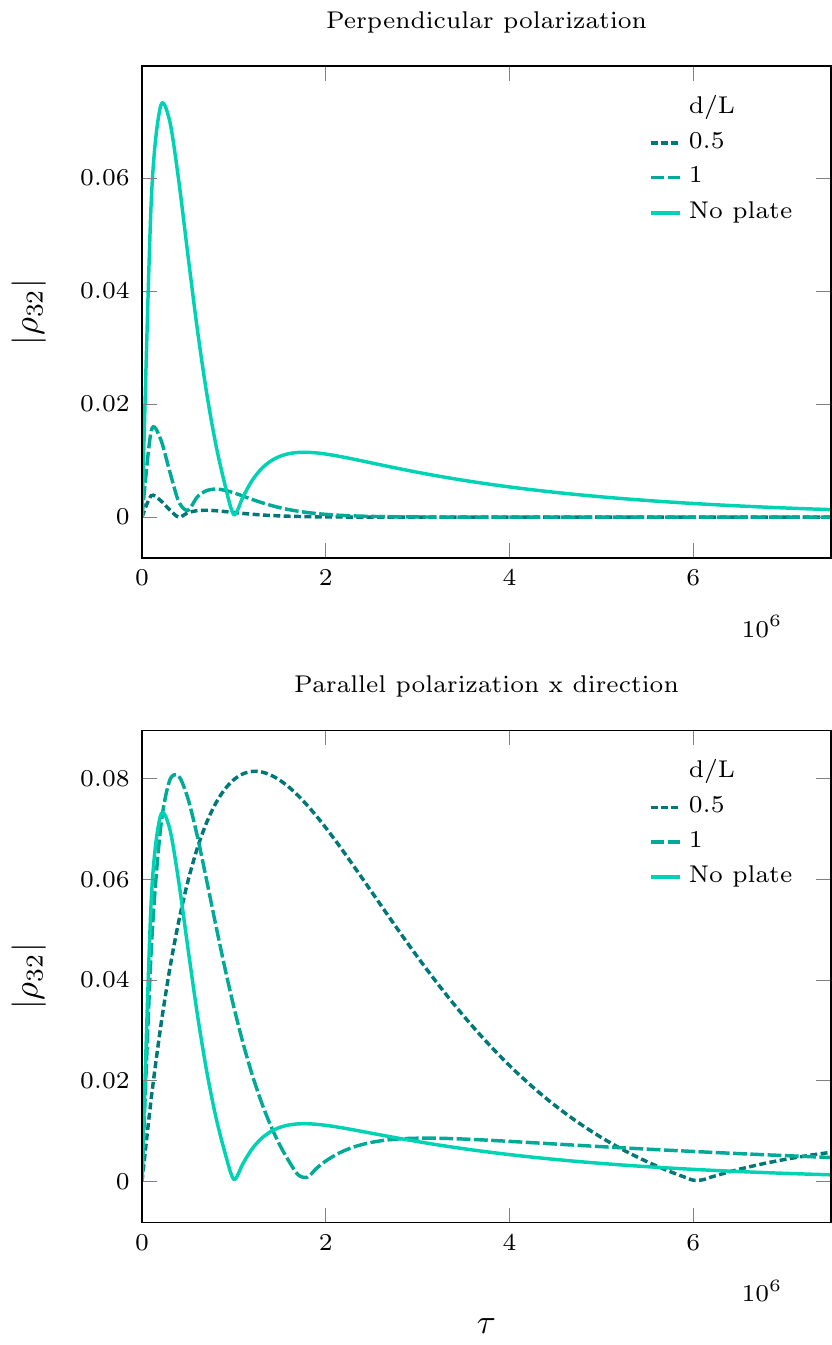}
\caption{\label{rho32}$\rho_{32}$ absolute value evolution. The solid line represents the behavior when the bipartite system is solely coupled to vacuum fluctuactions. Other lines account for the bipartite coupled to vacuum fluctuations at a fixed $d$ distance of a conducting plate: dotted line for $d/L=5$ while dashed line for $d/L=1$. The bipartite initial state  is a maximal entangled state ($p=1/2$). On top we consider both atoms with perpendicular dipole moment while at bottom, we represent the temporal evolution if both atoms have dipolar moments parallel to the plate.}
\end{figure}

\begin{figure}[ht]
\centering
\includegraphics[width=.8\columnwidth]{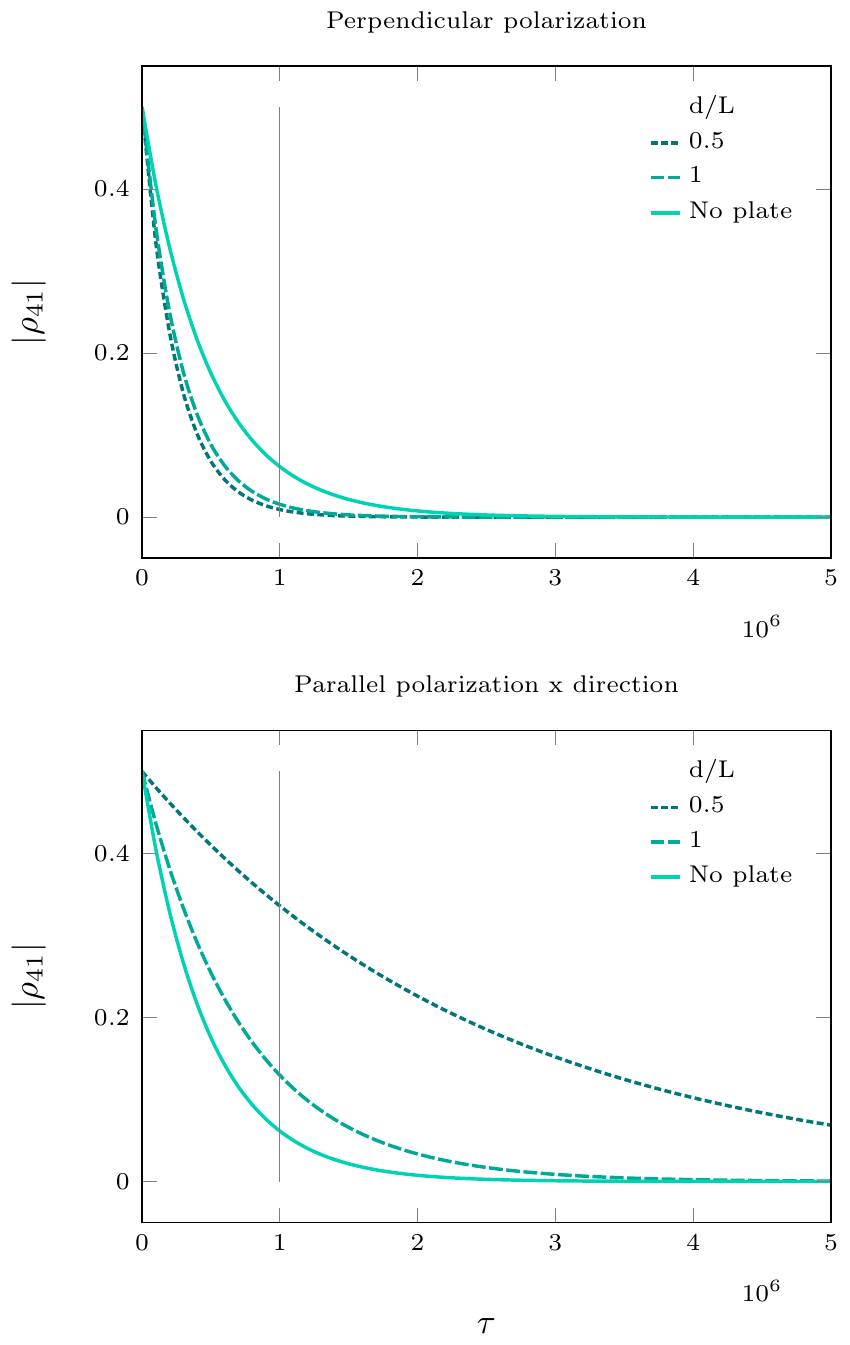}
\caption{\label{rho41} Evolution of $\rho_{41}$ absolute value. The solid line represents the behavior when the bipartite system is solely coupled to vacuum fluctuactions. Other lines account for the bipartite coupled to vacuum fluctuations at a fixed $d$ distance of a conducting plate: dotted line for $d/L=0.5$ while dashed line for $d/L=1$. The bipartite initial state  is a maximal entangled state ($p=1/2$). On top we consider both atoms with perpendicular dipole moment while at bottom, we represent the temporal evolution if both atoms have dipolar moments parallel to the plate.}
\end{figure}

In Figs. \ref{rho32} and \ref{rho41}, we show the evolution of the absolute value of the coherences with time. In each figure, we  compare the temporal evolution of the coherences of the reduced density matrix  when there is no plate present (solid line). In addition, we show the behavior of these quantities if there is a conducting boundary located at different distances (dotted and dashed lines) from the bipartite quantum system. In the latter case, we can further consider the orientation of the dipolar moment of the atoms. Hence, in both figures, we plot the temporal behavior for perpendicular orientation of both dipole moments on top, and parallel orientation of both atoms at the bottom.   As expected, both coherences  decay for sufficiently long times in all considered cases. Even yet, the effect introduced by the presence of the plate tends to vanish as the atoms are placed at larger distances, leaving decoherence effects solely to the zero point fluctuations of the electromagnetic field. 
However, we can note different decoherence timescales for the cases considered. We can surely define the decoherence timescale where interference terms vanish. As it can be seen in Fig.\ref{rho32}, vacuum fluctuations induced interference destruction around $\tau\sim 10^6$. If we consider this decoherence time $\tau_0$ as a reference timescale, we can note that decoherence occurs for shorter times when the dipole orientations are perpendicular to the plate ($\tau_{\perp}$). On the other hand, if orientations are parallel to the boundary, decoherence of the off-diagonal matrix elements takes longer $\tau_{\parallel}$. Therefore, we can state that $\tau_{\perp} < \tau_0 <\tau_{\parallel}$.
In addition,  we can further interpret this result in terms of the images method when the bipartite atom system is very close to the conducting surface. 
If the plane is parallel to the dipole, the image dipole is given by $\bf p_{\rm im} = -\bf p$. Therefore, the total dipole moment vanishes, and so does the probability to emit a photon. The image dipole cancels the effect of the real dipole and this produces less decoherence. On the other hand, when the conductor is perpendicular to the dipolar orientation, the image dipole is equal to the real dipole. Therefore, the total dipole is twice the original one. This in principle would lead us to conclude that the total decoherence factor grows \cite{villanueva}.

We can acquire a  better picture of the behavior of decoherence with the atom-plate distance by comparing the decoherence process in  free space (without plate) with that case in which the vacuum field state is modified by the presence of the  boundary condition imposed by the mirror. The results of this comparison, that has been separately  done for the elements $\rho_{32}$ and $\rho_{41}$.
\begin{figure}[h]
\centering
\includegraphics[width=.8\columnwidth]{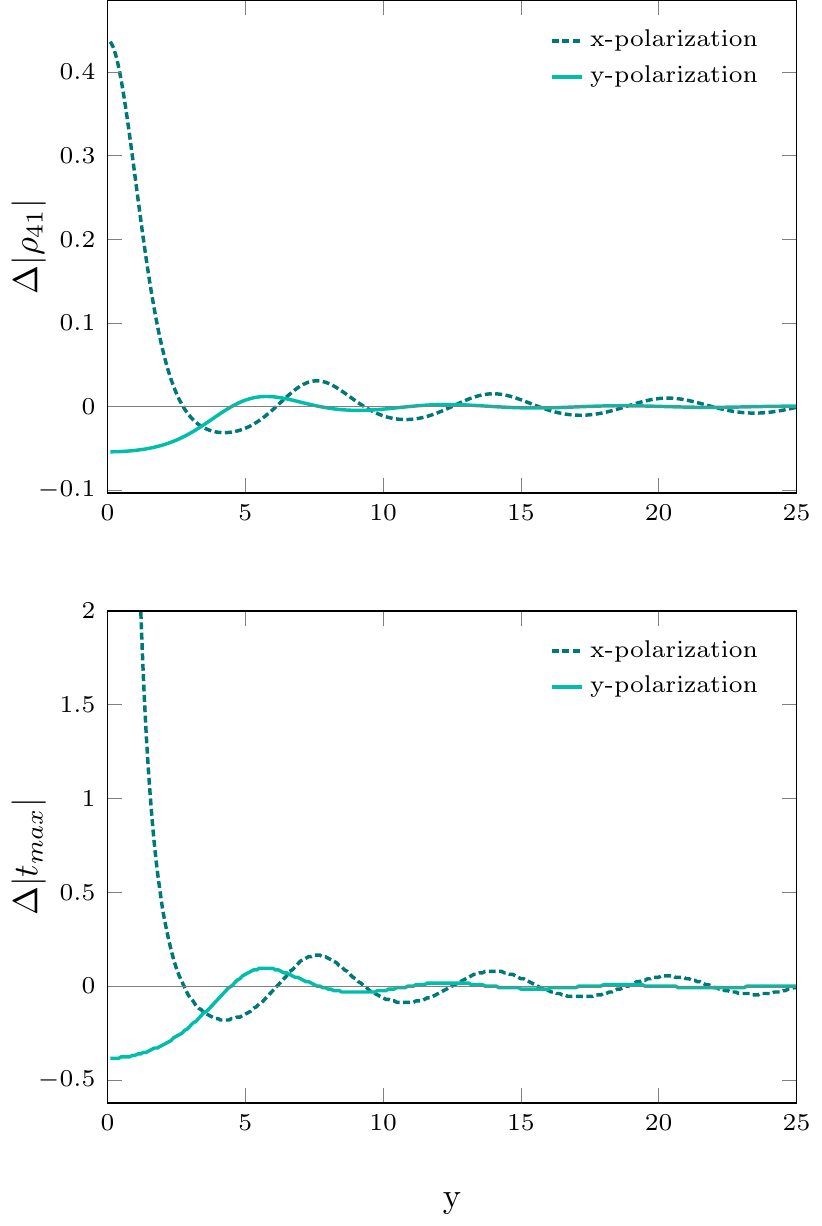}
\caption{\label{coherenciadiff} Comparison between the amplitude decay of the off-diagonal terms of the density matrix.}
\end{figure}
As it is suggested from the behavior of $|\rho_{41}|$ seen in Fig. \ref{rho41}, by taking the difference in this element's amplitude at a fixed time when the system is affected by free space electromagnetic field and by the electromagnetic field in presence of a perfectly conducting plate, we could acquire an idea of the relation between the effect of the environment in both cases. This difference 
\begin{equation}
     \Delta \abs{ \rho_{41}}= \left(\abs{\rho^{\text{plate}}_{41}}-\abs{\rho^{\text{free}}_{41}}\right)_{(t=\pi)}
     \label{delta_41}
\end{equation}
is illustrated in the upper figure of Fig, \ref{coherenciadiff} for both x and y polarized atoms. As it will be a recurrent behavior along this work, there is an initial regime in which the presence of the plate reinforces the effect of the environment if the atoms are polarized along the perpendicular y-direction and modifies this effect delaying the system behaviour when they are polarized along the parallel x-direction. This can be seen as, for a fixed time $t=\pi$ if the amplitude of the element has decreased more in presence of the plate than it would have in the free space case, then Eq. (\ref{delta_41}) should be negative, obtaining positive values of this subtraction in the opposite case (this fact can be seen for distances $d$ satisfying $d \leq 4 L$ approximately). The amplitude of this correction can be seen to decay with distance while oscillating and tends asymptotically to the free space case for distances $d \geq10 L$ approximately.
Considering the behavior shown by $|\rho_{32}|$  in Fig. \ref{rho32}, a subtraction similar to that performed for $|\rho_{41}|$ doesn't seem to be a good indicator of the modifications to the environment effect which are introduced by the presence of the plate. Instead, we have plotted the difference in time position of the local maximum present in the revival
\begin{equation}
    \Delta t_{\text{max}}=t_{\text{max}}^{\text{plate}}-t_{\text{max}}^{\text{free}}.
    \label{delta_32}
\end{equation}
The lower graph in (\ref{coherenciadiff}) consists on a plot of this difference $ \Delta t_{\text{max}}$ as a function of the distance from the particles to the plate.
We can see that in this case the modification induced by the plate exhibits the same qualitative behavior as that found for $|\rho_{41}|$.

\section{Entanglement dynamics}\label{concurrence}

Entanglement is one of the most intriguing properties of quantum mechanics, as it is a form of correlation that can not be explained in terms of any classical theory. Bipartite entanglement of pure states is conceptually well understood. When dealing with mixed states, we say that the state is entangled if it can not be written as a mixture of separable pure states. The dynamical behavior of correlations present in a composite open quantum system strongly depend on the noise produced by the surrounding environment. It is well known that quantum entanglement may be ruined due to the unavoidable interaction between the quantum system and its environment, which is one of the main challenges to the realization of quantum information technologies. In realistic situations the state of a quantum system is mixed. The so-called maximally entangled mixed states (MEMS) exhibits the maximum amount of entanglement for a given degree of mixedness. In this section, we shall study the entanglement dynamics of the bipartite system in the presence of vacuum fluctuations. We shall then compare the entanglement dynamics  to the one obtained if the bipartite system evolves in front of a conducting plate located at a fixed distance. In this last case considered, the dipole moments can be considered either perpendicular or parallel to the conducting plate.  Then, it might be easy to note if there exists a physical situation in which entanglement is robust or it may be generated as ``sudden birth phenomena " through the interaction of a common environment. Among the many physically motivated measures of entanglement for mixed states, entanglement of formation is intended to quantify the resources needed to create a given entangled state, but its exact computation involves a minimization over all possible pure-states decompositions which makes it inconvenient in the general case. However, in the particular case of a bipartite system consisting on two two-level subsystems, the quantity known as concurrence is monotonically related to entanglement of formation and, while its not so clearly motivated, can be taken as a measure of entanglement on its own \cite{concurrence}. The concurrence for the state described by $\rho_s(t)$ vanishes if 
$\rho_s(t)$ is a separable state and ranges monotonically to 1 for maximally entangled states. It can be computed as
\begin{equation}
    \mathcal{C}(\rho)=\text{max}(0, \sqrt{\lambda_1}-\sqrt{\lambda_2}-\sqrt{\lambda_3}-\sqrt{\lambda_4}),
\end{equation}
where $\lambda_i$ are the eigenvalues of $\Tilde{\rho}=\rho_s^*(\sigma^1\otimes\sigma^2)\rho_s(\sigma^1\otimes\sigma^2)$. 

\begin{figure}[ht]
\centering
\includegraphics[width=.8\columnwidth]{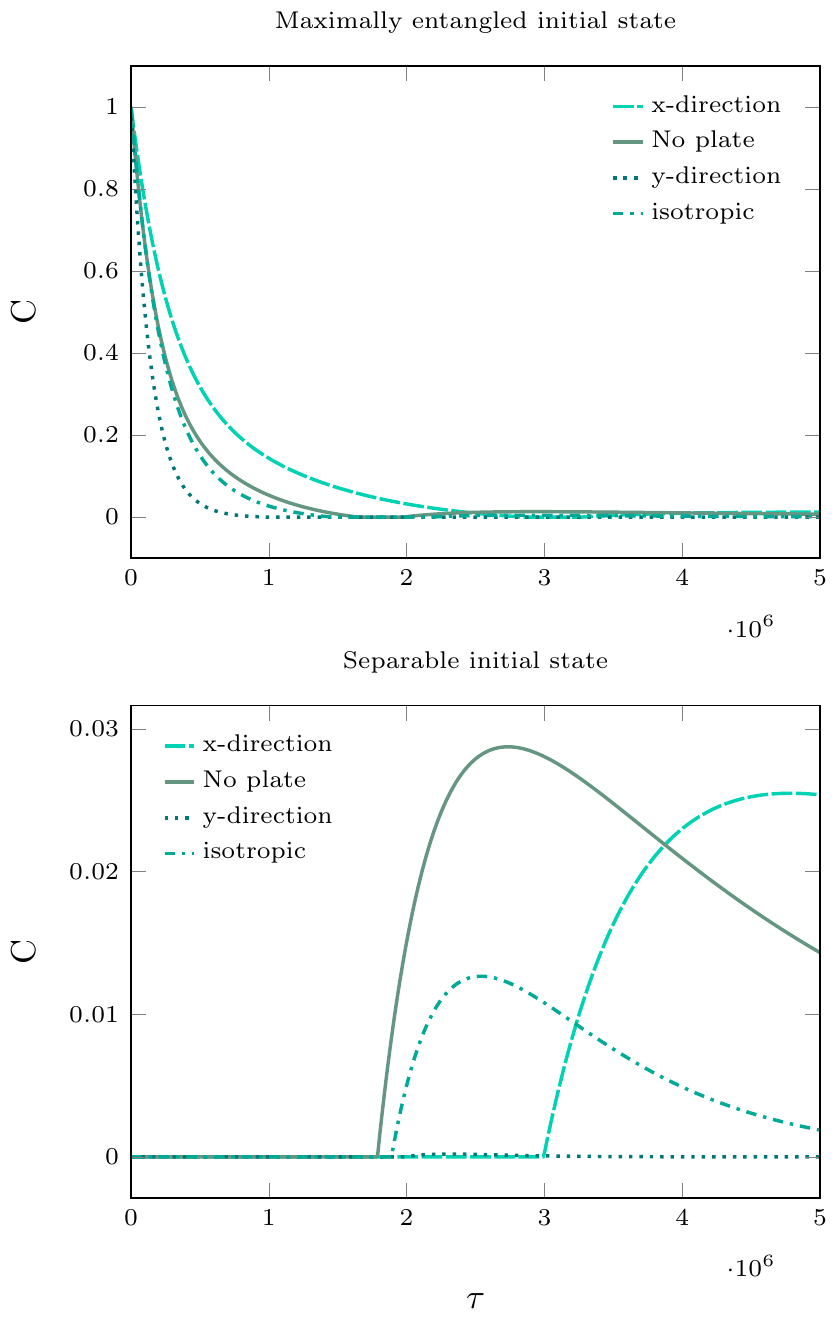}
\caption{Temporal evolution of concurrence for an initial maximal entangled state coupled to vacuum fluctuations (solid line)  and coupled to vacuum fluctuations at a fixed distance to a plate. Top: Dipole perpendicular orientation is represented with a dotted line while parallel dipole orientation with a dashed line. Bottom:  Parallel orientation is represented with a dotted line while parallel dipole orientation with a dashed line.}
\label{concurrence_p}
\end{figure}
\begin{figure}[ht]
\centering
\includegraphics[width=.8\columnwidth]{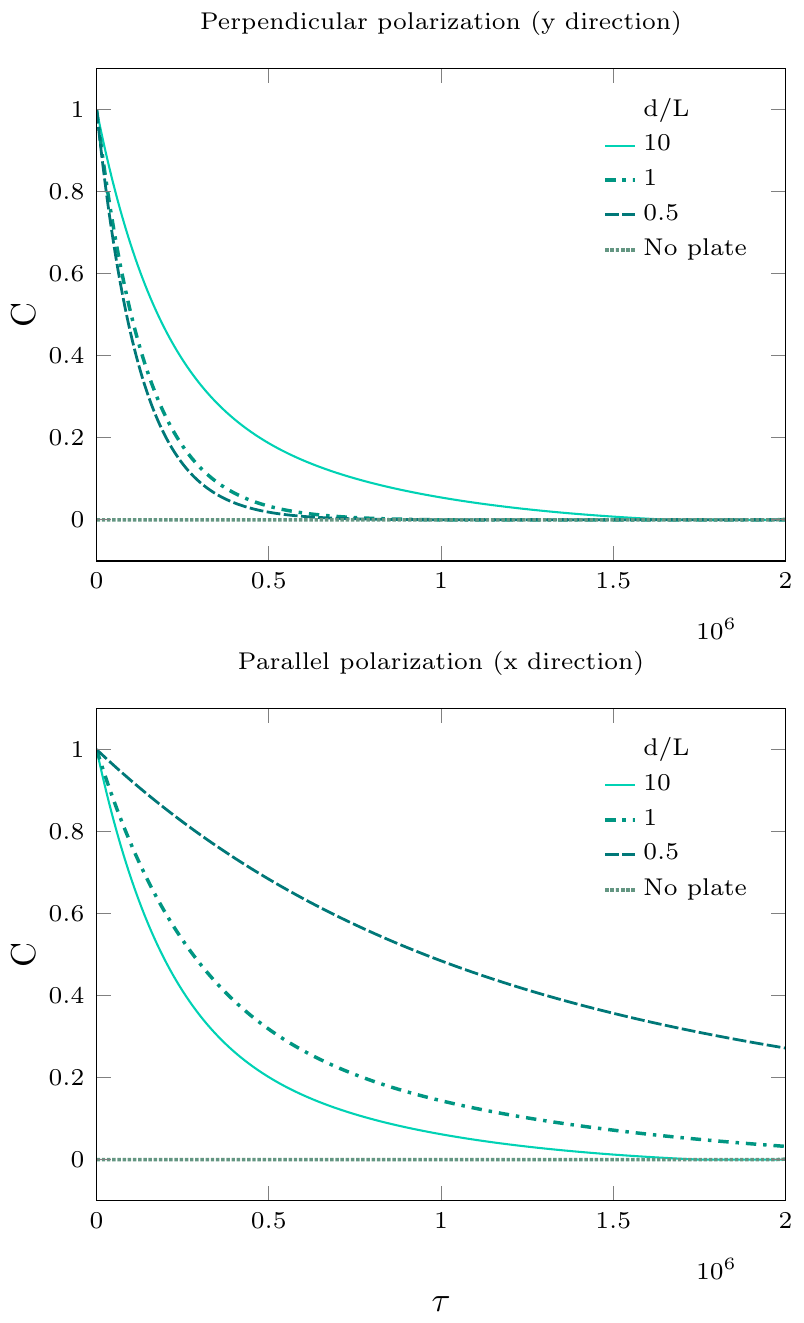}
\caption{Concurrence evolution for an initial maximally entangled state (p=0.5), considering the atoms fixed at different distances of the conducting plate. Top: perpendicular dipole orientation. Bottom:  parallel dipole orientation. }
\label{concurrence_y}
\end{figure}
In Fig.\ref{concurrence_p}, we show the temporal evolution of concurrence for a bipartite initial state of the form defined in Eq.(\ref{estado_1}). On top of the figure, we can see the concurrence for a MES for different situations considered: evolving under the presence of solely vacuum fluctuations (solid line), parallel dipole orientation (for each particle) evolving in front of a conducting plate (dashed line), perpendicular dipole orientation (dotted line) and an isotropic orientation (dotdashed line). Therein, we can note that concurrence tends to zero following the decoherence timescales hierarchy found before: entanglement of an initially entangled state is faster destroyed when the two atoms' dipole moments are perpendicular oriented to the plane compared to the presence of solely vacuum fluctuations and dipole moments orientations parallel to the plane. For some dipole orientations a revival of entanglement is seen, although its magnitude is much smaller than the initial value. At the bottom of Fig.(\ref{concurrence_p}), we present the temporal evolution of the concurrence induced by the environment in an initially separate bipartite state for the same situations considered on top. Therein, we can see that the strongest environment (dotted line) is not efficient enough to induce quantum entanglement (at least in the timescale shown). However, if a much weaker environment is considered, we can find that there is sudden generation of entanglement. It is easy to note, that the timescale at which entangled is created follows the inverse of the decoherence timescales hierarchy mentioned before. 
The existence of entanglement revivals and the entanglement generation should not be thought of as a sign of non-markovianity, as we have discarded all contributions from environmental memory effects when performing the markovian approximation. These features of entanglement dynamics are a consequence of the induced collective dynamics \cite{maniscalco2,ficek}. The radiated field produced by spontaneous emission of an atom influences the dynamics of the other atom through the vacuum field, but there is no information backflow from the field to the radiating atom.

In Fig. \ref{concurrence_y}, we present the temporal evolution of the concurrence for a MES  ($p=0.5$)  when the bipartite system evolves in front of a conducting plate, as function of the distance to the conducting plate $d$ for two different situations: (top) perpendicular-aligned atoms and (bottom) parallel-aligned atoms. We can see that the hierarchy is inverted among the different situations: for perpendicular orientations of the dipoles, the most destructive situation is nearer the plate.

\begin{figure}[ht]
\centering
\includegraphics[width=.8\columnwidth]{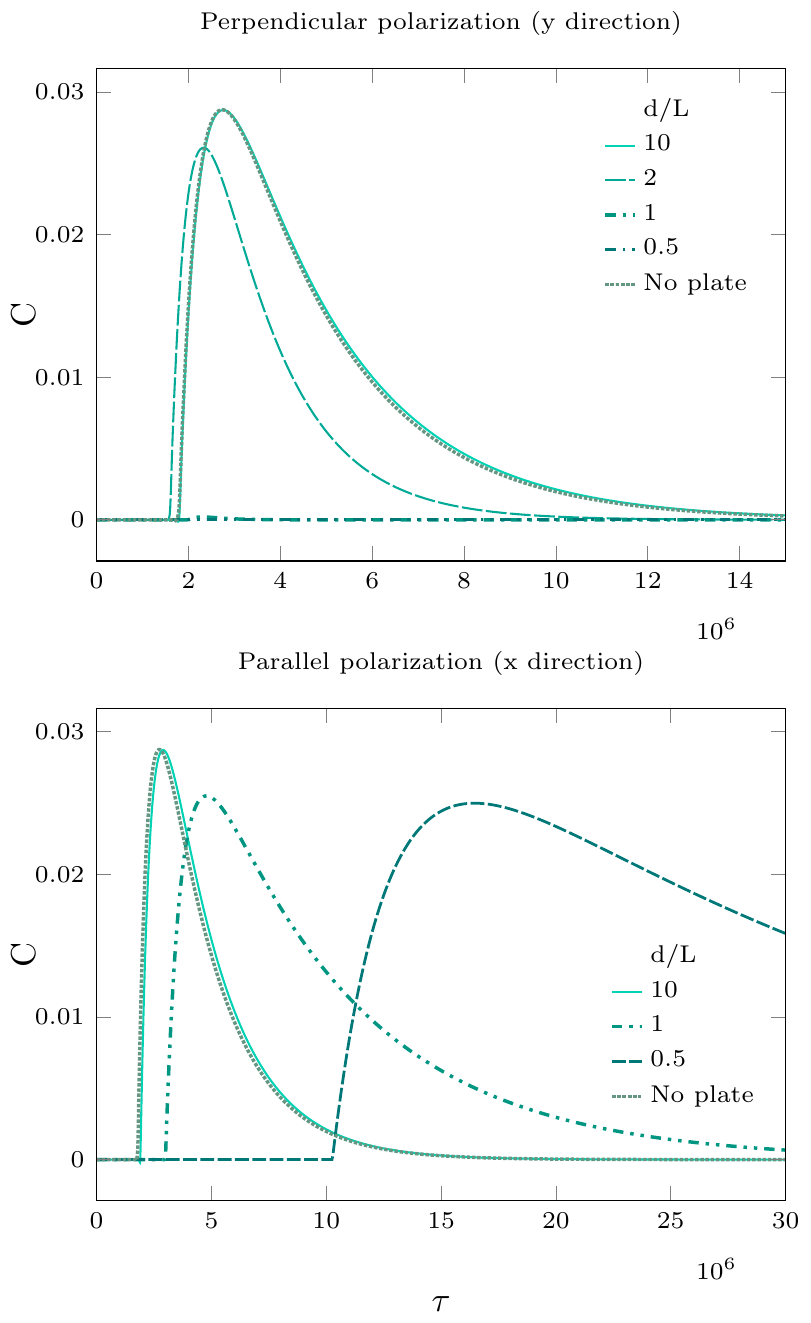}
\caption{Concurrence evolution (as a function of natural periods) for an initial separable state (p=1), considering the atoms fixed at different distances of the conducting plate. Top :perpendicular dipole orientation. Bottom:  parallel dipole orientation.}
\label{concurrence_y_2}
\end{figure}

In Fig.\ref{concurrence_y_2}, we plot the induced entanglement generation for an initially separate state for different distances of the bipartite to the plate. Therein, it is easy to note that entanglement birth is more likely to occur when the dipoles are oriented parallel to the conducting plate. On top of Fig. \ref{concurrence_y_2}, we can see that for short distances to the plate, no entanglement generation is induced when the dipoles are oriented perpendicular to the plate, in agreement with  the result obtained in Fig.\ref{concurrence_y}.
It can be seen that this correction tends to delay the revival for atoms polarized along the x-direction, while it is suppressed for atoms polarized along the perpendicular y-direction. By exploring the dependence of this correction with the distance from the atoms to the plate, as it is shown in Fig. \ref{concurrence_y}, we found that the revival is suppressed for atoms polarized along the perpendicular direction when they are placed close to the plate, appearing as the distance is increased.
When the dipolar coupling is along the x-direction, the presence of the plate seems to delay the appearance of the revival and extend its duration. Even when this behavior is not monotonic with distance, it can be seen that it is indeed monotonic for those distance intervals for which the correction is stronger.

In summary, we must note that the most decoherent situation for a MES state is when the bipartite system evolves in front of a conducting plate in the presence of vacuum fluctuations, with a perpendicular dipole moment orientation. In this case, the environment destroys initial quantum entanglement in a shorter time that if the evolution would have been done in free quantum vacuum fluctuations. This means that  in this particular situation the presence of a conducting boundary deteriorates quantum entanglement and does not help in the search of experimental situations that would preserve quantum entanglement. However, in the opposite situation, that is to say the bipartite system with parallel dipole orientation in front of a conducting plate, leads to an encouraging better situation: the decoherence time is delayed by the presence of a boundary and thus quantum entanglement is preserved longer.  If an experimental situation definitely contains a conducting plate, quantum entanglement is better preserved for a parallel dipole orientation at short distances  of the plate.

\section{Geometric phase} \label{phase}
In this section we shall study the geometric phase accumulated by the bipartite system when evolving under the presence of vacuum fluctuations. We shall
study the corrections to the unitary geometric phase acquired and compared it with that obtained when there is a conducting plate.
The geometric phase for a mixed state under non-unitary evolution has been defined  in the kinematic approach as \cite{tong1,tong2}

\begin{equation*}
     \phi_g=\arg\sum_{k}\sqrt{\epsilon_{k}(0)\epsilon_{k}(\tau)}\bra{\Psi_{k}(0)}\ket{\Psi_{k}(\tau)}e^{-\int_0^\tau dt \bra{\Psi_{k}}\ket{\Dot{\Psi}_{k}}},
\end{equation*}
where $\epsilon_k(t)$ are the eigenvalues and $\ket{\Psi(t)}$ are the eigenstates of $\rho_s(t)$. For a pure initial state, the expression above is simplified to

\begin{equation}
    \phi_g=\arg\bra{\Psi(0)}\ket{\Psi(\tau)}-\Im\int_0^\tau dt \bra{\Psi}\ket{\Dot{\Psi}},
    \label{fase_exacta}
\end{equation}
with $\ket{\Psi(t)}$ the eigenstate associated to the eigenvalue $\lambda(t)$ that satisfies $\lambda(0)=1$.
In the last definition, $\tau$ denotes a time after the total system completes
a cyclic evolution when it is isolated from the environment. 
Taking into account the effect of the environment, the system no longer undergoes a cyclic evolution. However, we shall consider a quasi cyclic path for a time interval $t \in [0,\tau]$ with $\tau=\pi/\omega_0$.
When the system is open, the geometric phase that would have been obtained if the system had been closed $\phi_u$ is modified. This means, in a general case, the phase is $\phi_g=\phi_u+\delta\phi$, where $\delta\phi$ is the correction to the unitary phase, induced by the presence of the environment (electromagnetic field and conducting plate) \cite{042311}. 
The eigenvalues of the reduced density matrix of the system can be found to be 
\begin{align*}
    \lambda^1_{\pm}= \frac{1}{2}\left(\rho_{11}+\rho_{44}\pm \sqrt{(\rho_{11}-\rho_{44})^2 + 4|\rho_{41}|^2}\right)\\
    \lambda^2_{\pm}= \frac{1}{2}\left(\rho_{22}+\rho_{33}\pm \sqrt{(\rho_{22}-\rho_{33})^2 + 4|\rho_{32}|^2}\right),
\end{align*}
and its easy to see that $\lambda^1_+(0)=1$ while the rest of the eigenvalues vanish at $t=0$. The eigenstate appearing in Eq.(\ref{fase_exacta}) is then
\begin{equation}
    \ket{\Psi}=\frac{-(\rho_{44}-\lambda^1_+ )\ket{11} + \rho_{41}\ket{00}}{\sqrt{(\rho_{44}-\lambda^1_+ )^2 + |\rho_{41}|^2}}.
\end{equation}
With this eigenstate, Eq.(\ref{fase_exacta}) reduces to the integral
\begin{equation}
    \phi_g=-(\omega_0 + c_{11} )\int_0^{\tau}\frac{|\rho_{41}|}{(\rho_{44}-\lambda^1_+)^2+|\rho_{41}|^2}.\label{exactGP} \\
\end{equation}
\begin{figure}[h]
\centering
\includegraphics[width=.9\columnwidth]{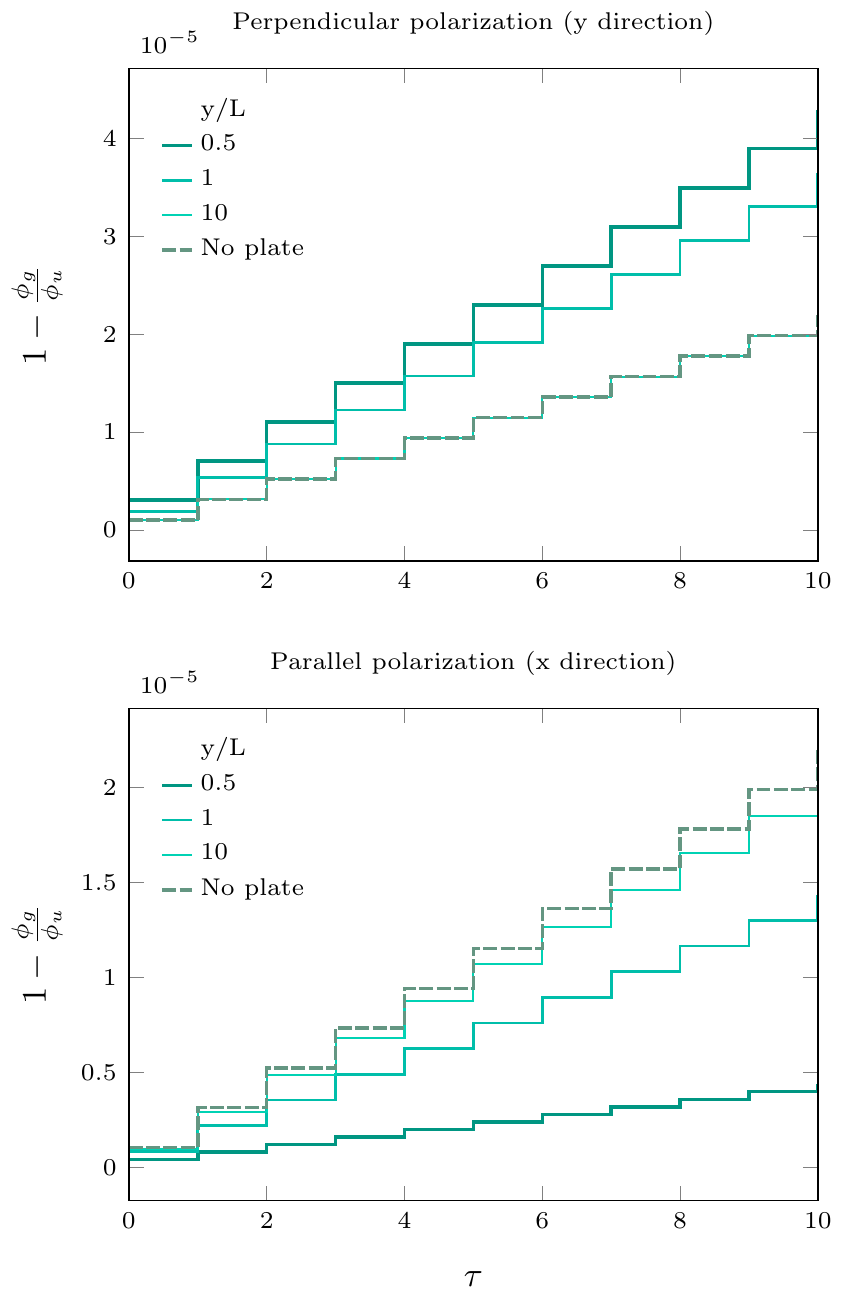}
\caption{Correction to the GP for different relations of the distance between atoms $L$ and the distance of the atoms to the plate $d$, for an initial state with p=0.5. Above: atoms polarized perpendicular to the plate. Bottom: atoms polarized parallel to the plate.}
\label{fase_p}
\end{figure}
We shall start by studying the geometric phase acquired under the influence of the environment. If the correction to the GP is negligible ($\delta \phi \ll \phi_u$), the GP acquired would be very similar to the unitary one $\phi_g \sim \phi_u$, and thus $1-\phi_g/\phi_u \approx 0$. If the correction becomes considerable, then this quantity would increase.
In Fig. \ref{fase_p}, we show the GP acquired by an initial maximally entangled state for different winding numbers $N=t/\tau$. On top, we can find the situation where the dipoles are perpendicular-oriented while at bottom, the dipoles are oriented parallel to the plate. In both cases, we show how the correction to the GP is modified as the bipartite system is located at different distances from the plate $d/L$ and compare each case to the correction obtain when the system is only coupled to vacuum fluctuations (solid line). If we choose a fixed time or cicle, i.e. $N=5$, we can see that the correction to the GP is smaller for small distances to the conducting plate when the dipoles are oriented parallel to the surface. On the contrary, if the dipoles are perpendicular to the surface, the smallest correction belongs to a fixed distance far enough from the plate (as if there were no plate at all). 
The correction to the GP tends in most cases to be enlarged when compared to the correction induced in the free (no mirror) space when the particles are polarized in the ${\check y}$-direction and to be diminished when they are polarized in the parallel ${\check x}$-direction, in accordance with the results obtained for the coherences and the concurrence of the state. As expected, these corrections tend to the free space correction for big enough distances in all cases. \\

\begin{figure}[ht]
\centering
\includegraphics[width=.8\columnwidth]{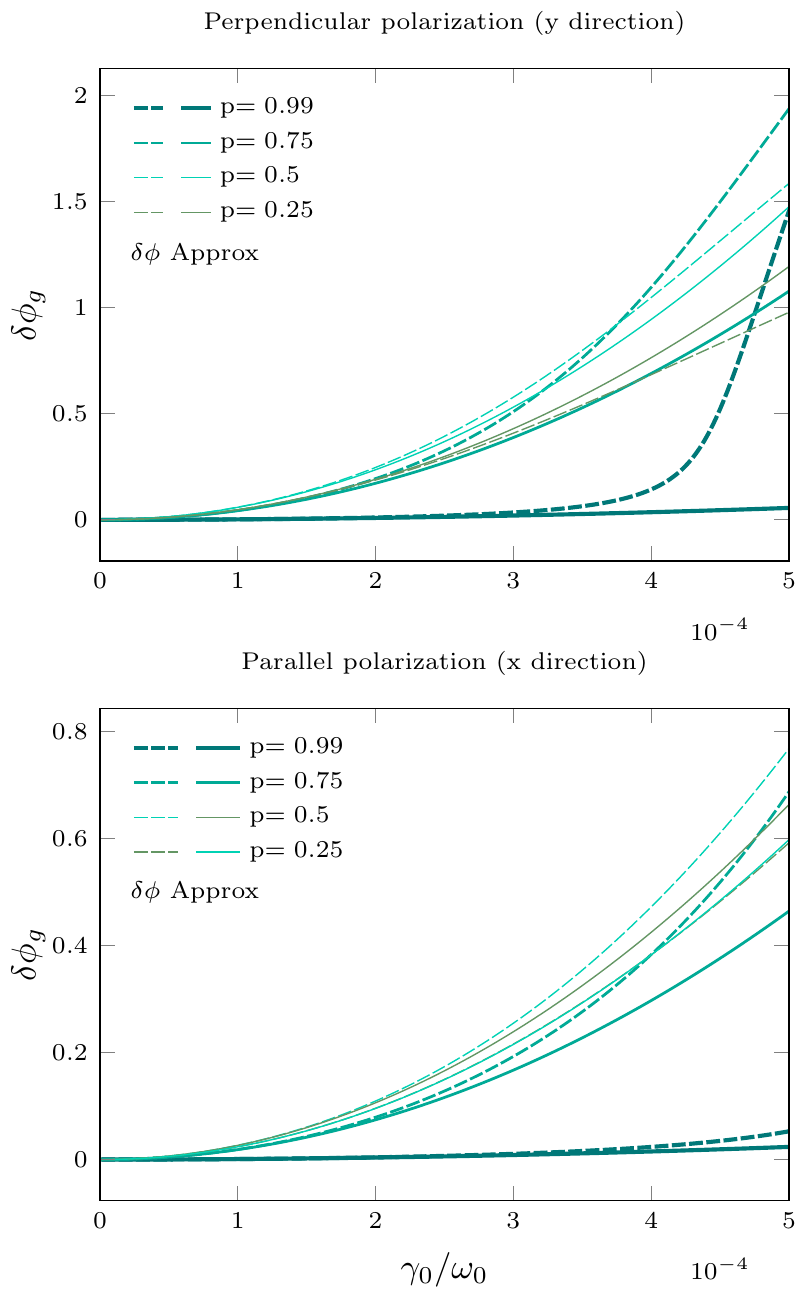}
\caption{Exact GP correction compared to first order approximation for different initial states. Top: perpendicular-oriented dipoles. Bottom: parallel-oriented dipoles.}
\label{delta_fase}
\end{figure}

In order to obtain an analytical expression of the correction $\delta\phi$ and get an insight into its functional dependence, 
we can further perform an expansion of this phase for small $\gamma_0/\omega_0$ (weak coupling limit) up to second order, 
\begin{align}
    &\phi_g \approx -2\pi (1-p)\left[1 + (c_{11}+2\pi p a_{11})\frac{1}{\omega_0} + \right.    \label{fase_aprox} \\[.8em] \nonumber
    &\left.+ \frac{2\pi p}{3\omega_0^2}\left[3a_{11}c_{11}+4\pi p a_{12}^2-4\pi(1-3p)a_{11}^2\right]\right] +\mathcal{O}\left(\frac{\gamma_0}{\omega_0}\right)^3.
\end{align}
The first term in the above expansion is $\phi_u = -2\pi(1-p)$, the unitary geometric phase (this is the phase we would have obtained when the system is isolated). We can see that the first order correction depends only on $a_{11}$ and $c_{11}$ meaning the correction is, to this order, independent of the separation between the atoms $L$, which contributes only to the following order. However it is not independent of the fact that there are two particles, as it is reflected in the quadratic dependence of the first order correction, compared to the cubic dependence found in works previously done \cite{fasechinos,lombardovillar022115}.
Further analysis can be done over the phase if we write the expressions for $a_{11}$ and $c_{11}$ so that the correction, to first order in $\gamma_0/\omega_0$ reads
\begin{eqnarray}
    \delta\phi \approx &-& 4\pi^2 (1-p) p  \frac{\gamma_0}{\omega_0} \Big[1 -       \nonumber \\ 
     &- & 3 \sum_{m=1}^3(r^1_m)^2  \left( b^{11}(d,\omega_0) - \frac{h^{11}(d,\omega_0)}{2\pi^2 p} \right) \Big], 
    \label{correccion_fase}
\end{eqnarray}
where it becomes clear that the correction has a free-space component $-4\pi^2(1-p)p$ and a contribution induced by the presence of the plate. The correction to the phase obtained in Eq.(\ref{correccion_fase}) has an strong dependence on the distance to the plate at which the atoms are fixed due to the $1/y^3$ dependence of the frequency shift $h^{11}$. For distances in the considered range the correction takes its maximum value at
\begin{equation}
    p=\frac{1}{2}-\frac{3\sum_{m=1}^3(r^1_m)^2h^{11}(d,\omega_0)}{4 \pi^2  \left(1-3\sum_{m=1}^3(r^1_m)^2b^{11}(d,\omega_0)\right)}.
    \label{correccion_max}
\end{equation}
for any polarization. For this last particular case (\ref{correccion_max}) there is a maximum correction to the phase given by 
\begin{equation*}
\delta\phi_{\text{max}}=\frac{(2 \pi  a_{11}+c_{11})^2}{4 a_{11}\omega_0 }.
\end{equation*}
It is clear that the maximum correction to the phase (to first order in $\gamma_0/\omega_0$) occurs for the maximally entangled state in the free space case and it is modified by the presence of the conducting plate. 
In order to get a picture of the range of validity of this approximation we have studied the variation of both the exact phase difference and the first order approximation as a function of the expansion parameter $\gamma_0/\omega_0$, as shown in Fig. \ref{delta_fase}. We did so for different initial states and for both studied polarization directions of the atoms, considering they are fixed at a distance $d=L$ from the plate.

For most studied p-values the first order approximation faithfully reproduces the exact behavior for $\gamma_0/\omega_0$ values ranging up to $10^{-4}$, starting to be distinguishable for $\gamma_0/\omega_0$ values of order $2. 10^{-4}$. This range of validity seems to be similar for both polarization directions. The specific case of $p=0.99$, which represents an initial state very close to $\ket{11}$ exhibits a different behavior as the approximation remains valid for $\gamma_0/\omega_0$ values up to three times bigger.


\section{Conclusions}\label{conclusiones}

In this article, we have studied the complete dynamics of a bipartite two-level state system (atoms) coupled to an electromagnetic field. We have derived the complete master equation and obtained the reduced density matrix. When computing  the environmental kernels for the vacuum fluctuations, we further contemplate the situation of having a conducting plate at a fixed distance $d$ of the bipartite system. In this way, we have obtained environmental kernels comprised of terms depending solely upon vacuum fluctuations and other terms with a clear dependence upon the distance to the plate.  Once we have obtained the reduced density matrix for a general state of the bipartite system, we studied the dynamics of a Bell-like initial state. We have defined the decoherence timescale for vacuum fluctuations and compared the decoherence timescale when there is a conducting plate. As for the latter situation, we defined two clearly situations: (i) atomic dipole orientations parallel to the boundary and (ii) atomic dipole orientations perpendicular to the plate. We have obtained that there is a hierarchy in the decoherence timescales for the different situations considered. This result could be easily understood in terms of the image method, concluding that having a dipole orientated perpendicular to the plate, can be demonstrated to originate a more noisy environment and therefore lead to a decoherence timescale shorter than in the case of the dipole is parallel oriented. 

Further, we have studied the temporal evolution of concurrence for the initial bipartite state, starting with an entangled initial state. We have also analyzed the creation of entanglement in a separable initial state due to the interaction with the environment. We have found that entanglement is most likely to be created for an initially separate state if the dipoles are oriented parallel to the surface.

Finally, we have computed the accumulated geometric phase acquired  by a bipartite system in the presence of the external environment.  
We have considered an initial maximally entangled state and studied how the geometric phase is corrected for each case considered, and compared the results to the correction obtained if the bipartite have evolved in free space (only vacuum fluctuations).  We have further considered how the GP is corrected if the bipartite state is located at different distances to the plate, finding that it is most corrected for small distances to the plate in the case the dipoles are oriented perpendicular to the boundary. 
We have also performed an analytical expansion in order to determine the different contributions to the correction of the geometric phase. We have found two types of contribution to the correction of the GP: (i) a contribution induced by vacuum fluctuations and (ii) a contribution induced by the presence of a conducting boundary. This interesting result reinforces the idea that the geometric phase has become a fruitful venue to explore indirectly quantum properties of a system with the emergence of new technologies.
 
All in all, we have presented a model in which we can exploit the quantum vacuum structure rendering a good scenario
for measurements of the geometric phase. It  has been argued that the observation of GPs should be done in times long enough to obey the adiabatic approximation but
short enough to prevent decoherence from deleting all phase information. This means that while there are dissipative and diffusive effects that induce a correction
to the unitary GP, the system maintains its purity for several cycles, which allows the GP to be observed. Particularly, we have shown that if we want to take advantage of the boundary induced structure modification of quantum vacuum, we shall explore an experimental setup at short distance of a reflecting mirror with a bipartite system composed of parallel oriented dipoles.  In such a situation, we shall find that the geometric phase acquired is similar to the unitary one at short times.
 \\
\section*{Acknowledgements}

This work was supported by ANPCyT, CONICET, and Universidad de Buenos Aires; Argentina. \\
FCL acknowledges ICTP-Trieste ans Simons Associate Program
\appendix
\section{Environment kernels }\label{appendixa}
The environment $a_{ij}(t)$ and $c_{ij}(t)$ kernels, are defined as:
\begin{align}
    a_{ii}&=\gamma_0\sum_{m=1}^3(r^i_m)^2\left[f^{ii}(\omega_0)-3b^{ii}(d,\omega_0)\right]\\\nonumber
    c_{ii}&=-\frac{3\gamma_0}{\pi}\sum_{m=1}^3(r^i_m)^2h^{ii}(d,\omega_0)\\\nonumber
    a_{12}&=a_{21}=3\gamma_0\sum_{m=1}^3 r_m^1 r_m^2\left[f^{12}(L,\omega_0)-b^{12}(L,d)\right]\\\nonumber
    c_{12}&=c_{21}=3\gamma_0\sum_{m=1}^3 r_m^1 r_m^2\left[g^{12}(L,\omega_0)-h^{12}(L,d,\omega_0)\right].
\end{align}
In the above set of equations,  $\gamma_0=\frac{|\mathbf{r}|^2\omega_0^3}{3}$, remembering we're working in natural units $c=\hbar=1$. Therein, we can identify different contributions of the environment: 
\begin{description}
\item (i) terms derived from solely vacuum fluctuations, such as $f^{ii}$ which does not depend on any distance and represent the effect of the environment as generator of spontaneous emission process.
\item (ii) terms derived from solely vacuum fluctuations, such as  $f^{ij}$ and $g^{ij}$, expressed in terms of an adimensional variable $x=L\omega_0$, where $L$ is the distance among quantum particles and represent an effective influence of each particle on the other due to their coupling with the electromagnetic field. This influence manifests itself both in an effective dipole-dipole interaction and in the collective damping factor.
\item (iii) terms comprising the presence of the reflecting plate, such as $b^{ij}$ and $h^{ij}$, expressed in terms of adimensional variables $y=2d\omega_0$ and $z=\sqrt{x^2+y^2}$, where $d$ is the distance of both particles to the plate. Those terms depending on $y=2d$ can be thought as effective actions exerted on an atom by its image dipole while those depending on $z=\sqrt{x^2+y^2}$, as effective actions exerted on an atom by the image dipole of the other atom. This can be seen from Im.\ref{esquema} where it is clear that those are the distances between the referred particles.
\end{description}

The explicit form of these contributions is:
\begin{align*}
    f^{ii}&=1\\
    b^{ii}(y)&=\frac{\delta_{m1}+\delta_{m3}}{2}\frac{A(y)}{y^3}-\delta_{m2}\frac{B(y)}{y^3}\\
    h^{ii}(y)&=\frac{\delta_{m1}+\delta_{m3}}{2y^3}\left[\,y - \ci{y}A(y)-\si{y}\Tilde{A}(y)\right] \\&-\frac{\delta_{m2}}{y^3}\left[ \ci{y}B(y)-\si{y}\Tilde{B}(y)\right]\\
    f^{12}(x)&=\frac{\delta_{m1}+\delta_{m2}}{2}\frac{A(x)}{x^3}-\delta_{m3}\frac{B(x)}{x^3}\\
    b^{12}(z)&=\frac{\delta_{m1}}{2}\frac{A\left(z\right)}{z^3}+\delta_{m3}\frac{C\left(z\right)}{z^3}-\frac{\delta_{m2}}{2}\frac{D\left(z\right)}{z^3}\\
    g^{12}(x)&= \frac{\delta_{m1}+\delta_{m2}}{2} \frac{\Tilde{A}(x)}{x^3} + \delta_{m3}\frac{\Tilde{B}(x)}{x^3} \\
    h^{12}(z)&= \frac{\delta_{m1}}{2}\frac{ \Tilde{A}\left(z\right)}{z^3}+ \frac{\delta_{m2}}{2} \frac{\Tilde{D}\left(z\right)}{z^5} + \delta_{m3}\frac{\Tilde{C}\left(z\right)}{z^5} ,
\end{align*}

among the definition of the functions appearing therein:
\begin{align*}
A(x)&=x\cos(x)+(x^2-1)\sin(x)\\
\Tilde{A}(x)&=x\sin(x)-(x^2-1)\cos(x)
\end{align*}
\begin{align*}
B(x)&=x\cos(x)-\sin(x)\\
\Tilde{B}(x)&=x\sin(x)+\cos(x)\\
C(u)&=\left[\left(\frac{y^2}{2}-z^2\right)u\cos(u)+\left(x^2+\frac{y^2}{2}(u^2-1)\right)\sin(u)\right]\\
\Tilde{C}(u)&=\left[\left(\frac{y^2}{2}-z^2\right)u\sin(u)-\left(x^2+\frac{y^2}{2}(u^2-1)\right)\cos(u)\right]\\
D(u)&=\left[\left(x^2- 2y^2\right) u\cos(u)+\left(2y^2+x^2(u^2-1)\right)\sin(u)\right]\\
\Tilde{D}(u)&=\left[\left(x^2- 2y^2\right) u\sin(u)-\left(2y^2+x^2(u^2-1)\right)\cos(u)\right]
\end{align*}

One can easily verify that  the corrections $b^{ij}$ and $h^{ij}$ induced by the presence of the conducting plate vanish for long enough distances of the particles to it, as they behave as inverse powers of this distance.

\section{Reduced density matrix elements}\label{appendixb}
Herein, we show the analytic expression of the components of the reduced density matrix elements, after having solved the master equation Eq.(\ref{eqcorta}) 
by assuming an initial quantum state of the form $\ket{\psi}=\alpha\ket{11}+\beta\ket{10}+\gamma\ket{01}+\sigma\ket{00}$:
\begin{widetext}
\vspace{-.75cm}
\begin{align}\nonumber
\rho_{11}(t)&=\alpha^2 e^{-4\Gamma^{11}}\\ \nonumber
 \rho_{41}(t)&=\sigma^*\alpha e^{-2\Gamma^{11}}e^{2i(\gamma^{11}+\omega_0 t)}\\ \nonumber
  \left(\begin{array}{c}
             \rho_{21}  \\
             \rho_{31}
        \end{array}\right)&=\left[\frac{ (\beta+\delta)\alpha}{2} \left(\begin{array}{c}
             1  \\
             1
        \end{array}\right) e^{-\Gamma^{12}}e^{-i\gamma^{12}}+ \frac{ (\beta-\delta)\alpha}{2}\left(\begin{array}{c}
             1  \\
             -1
        \end{array}\right)e^{\Gamma^{12}}e^{i\gamma^{12}} \right]e^{-3\Gamma^{11}}e^{i(\gamma^{11}+\omega_0t)}\\\nonumber
         \left(\begin{array}{c}
             \rho_{42}  \\
             \rho_{43}
        \end{array}\right)&=\left[\frac{(\sigma+\alpha \;G'(t))(\beta+ \delta)}{2} \left(\begin{array}{c}
             1  \\
             1
        \end{array}\right) e^{-\Gamma^{12}}e^{i\gamma^{12}} +\frac{(\sigma+\alpha \;F'(t))(\beta- \delta)}{2}\left(\begin{array}{c}
             1  \\
             -1
        \end{array}\right)e^{\Gamma^{12}}e^{-i\gamma^{12}} \right]e^{-\Gamma^{11}}e^{i(\gamma^{11}+2\omega_0t)}\\\nonumber
        \left(\begin{array}{c}
             \rho_{22}  \\
             \rho_{23}  \\
             \rho_{32}  \\
             \rho_{33}
        \end{array}\right)&=\left[\left( \alpha^2 F(t)+\frac{(\beta-\delta)^2}{4}\right) \left(\begin{array}{c}
             1  \\
            - 1  \\
            - 1  \\
             1
        \end{array}\right) e^{2\Gamma^{12}} +\left( \alpha^2 G(t)+\frac{(\beta+\delta)^2}{4}\right)\left(\begin{array}{c}
             1  \\
             1  \\
             1  \\
             1
        \end{array}\right)e^{-2\Gamma^{12}} \right] e^{-2\Gamma^{11}} + \\
        & \hspace{.5cm}+ \frac{-\beta^2+\delta^2}{2}\left(\begin{array}{c}
             -\cos2\gamma^{12}  \\
             -i\sin2\gamma^{12}  \\
             i\sin2\gamma^{12} \\
             \cos2\gamma^{12}
        \end{array}\right)
        e^{-2\Gamma^{11}},
\end{align}
\vspace{-.75cm}
\end{widetext}
The time dependent factors in these expressions are obtained by tracing out the environment 
\begin{align}
    \Gamma^{ij}&=\int_0^t dt' a_{ij}(t'),\nonumber \\
    \gamma^{ij}&=\int_0^t dt' c_{ij}(t'),\\\nonumber
     F'(t)&=\int_0^t dt'\,\left[a_{11}(t')-a_{12}(t')\right]e^{-2(\Gamma^{11}-i\gamma^{12}),}\nonumber
 \end{align} 
 
 \begin{align}
    G'(t)&=\int_0^t dt'\,\left[a_{11}(t')+a_{12}(t')\right]e^{-2(\Gamma^{11}+i\gamma^{12}),}\nonumber\\
    F(t)&=\int_0^t dt'\,\left[a_{11}(t')-a_{12}(t')\right]e^{-2(\Gamma^{11}+\Gamma^{12}),}\\\nonumber
    G(t)&=\int_0^t dt'\,\left[a_{11}(t')+a_{12}(t')\right]e^{-2(\Gamma^{11}-\Gamma^{12}),}.\nonumber
\end{align}

\bibliographystyle{apsrev4-1}
\bibliography{twolevels.bib}




\end{document}